\documentclass[a4paper, amsfonts, amssymb, amsmath, reprint]{revtex4-2}

\newcommand{\mywidth}{1.0}

\usepackage[english]{babel}
\usepackage[utf8]{inputenc}
\usepackage[T1]{fontenc}
\usepackage{graphicx,import}
\usepackage{units}
\usepackage{layouts}
\usepackage{setspace}

\usepackage[pdftex, pdftitle={Article}, pdfauthor={Author}]{hyperref} 

\renewcommand{\rm}[1]{\mathrm{#1}}

\def\ddt{\frac{\mathrm{d}}{\mathrm{d}t}}

\renewcommand{\u}[1]{\,\mathrm{#1}}

\DeclareMathOperator{\sech}{sech}

\usepackage{xcolor}

\usepackage[capitalize]{cleveref}

\DeclareMathOperator*{\argmin}{arg\,min}
\DeclareMathOperator*{\argmax}{arg\,max}
\newcommand{\norm}[1]{\left\lVert#1\right\rVert}

\usepackage{tikz}
\usepackage{tikzscale}
\usetikzlibrary{matrix}
\usetikzlibrary{fit}
\usetikzlibrary{backgrounds}
\usetikzlibrary{positioning}
\usetikzlibrary{arrows}
\usetikzlibrary{shapes}

\begin{document}

\title{Data-Driven Acceleration of Multi-Physics Simulations}

\author{Stefan Meinecke}
    \email[Correspondence email address: ]{meinecke@tu-berlin.de}
    \affiliation{Institut für Theoretische Physik, Technische Universität Berlin, Hardenbergstr. 36, 10623 Berlin, Germany}

\author{Malte Selig}
    \affiliation{Institut für Theoretische Physik, Technische Universität Berlin, Hardenbergstr. 36, 10623 Berlin, Germany}

\author{Felix Köster}
    \affiliation{Institut für Theoretische Physik, Technische Universität Berlin, Hardenbergstr. 36, 10623 Berlin, Germany}

\author{Andreas Knorr}
    \affiliation{Institut für Theoretische Physik, Technische Universität Berlin, Hardenbergstr. 36, 10623 Berlin, Germany}

\author{Kathy Lüdge}
    \email[Correspondence email address: ]{kathy.luedge@tu-ilmenau.de}
    \affiliation{Technische Universität Ilmenau, Institut für Physik, Weimarer Straße 25, 98693 Ilmenau, Germany}

\begin{abstract}
Multi-physics simulations play a crucial role in understanding complex systems. However, their computational demands are often prohibitive due to high dimensionality and complex interactions, such that actual calculations often rely on approximations. To address this, we introduce a data-driven approach to approximate interactions among degrees of freedom of no direct interest and thus significantly reduce computational costs. Focusing on a semiconductor laser as a case study, we demonstrate the superiority of this method over traditional analytical approximations in both accuracy and efficiency. 
Our approach streamlines simulations, offering promise for complex multi-physics systems, especially for scenarios requiring a large number of individual simulations.
\end{abstract}


\maketitle

\section{Introduction} \label{sec:intro}
Multi-physics simulations have become a powerful tool for analyzing and understanding complex physical systems. These simulations involve the integration of multiple physical models that describe the behavior of different phenomena, such as heat transfer, fluid dynamics, and electromagnetic fields. Laser simulations are a prime example of multi-physics simulation, since they are used to model the behavior of lasers in various applications, such as material processing \cite{gu2021material}, medical treatments \cite{LOE96,JUH99,NAG09}, and communication systems \cite{KNO00, KUN07a, RAF11}. However, performing these simulations can be computationally intensive and time-consuming \cite{CHO99, LIN13, KIL14, STR14a}. 

The computational costs are typically driven by a large number of degrees of freedom, which are subject to non-trivial, i.e., nonlinear mutual interactions \cite{STR14a}. 
In many cases, however, one is only interested in the dynamics of only a few of them. 
Due to their interdependence, one nonetheless needs to solve the complete system, to obtain the dynamics of interest. 

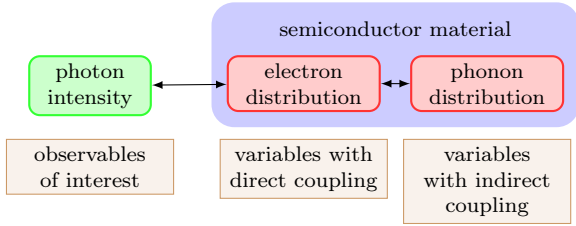
\begin{figure}[htbp]
\centering
    
\tikzstyle{semic_state}=[rectangle,
    thick,
    minimum height=0.6cm,
    minimum width=1.0cm,
    text width=1.8cm, 
    text centered,
    draw=red!80,
    fill=red!20,
    rounded corners]

\tikzstyle{phot_state}=[rectangle,
    thick,
    minimum height=0.6cm,
    minimum width=1.0cm,
    text width=1.4cm, 
    text centered,
    draw=green!80,
    fill=green!20,
    rounded corners]

\tikzstyle{dummy}=[rectangle,
    minimum height=0.6cm,
    minimum width=0.1cm]

\tikzstyle{desc}=[rectangle,
    minimum height=0.6cm,
    minimum width=0.6cm,
    text width=2.0cm,
    text centered,
    fill=brown!10,
    draw=brown!80,]

\tikzstyle{sdesc}=[rectangle,
    minimum height=0.6cm,
    minimum width=0.6cm,
    text centered,
    fill=blue!20,
    draw=blue!80,]

\tikzstyle{semic_background}=[rectangle,
    fill=blue!20,
    inner sep=0.2cm,
    rounded corners=3mm]

\begin{tikzpicture}[>=latex]
\footnotesize
\setstretch{1.0}

  \matrix (mtrx) [row sep=0.3cm, column sep=0.2cm, matrix of nodes, nodes in empty cells] {
    \node (PT) [phot_state]{photon intensity};  &  & \node (EL) [semic_state]{electron distribution}; & \node (PN) [semic_state]{phonon distribution};\\
    \node (desc1) [desc] {observables of interest}; & & \node (desc2) [desc] {variables with direct coupling}; & \node (desc1) [desc] {variables with indirect coupling}; \\
    };

    \node (sclabelanch) [fit=(EL) (PN)]{};
    \node[above=0.1cm of sclabelanch, inner sep=0.5mm, outer sep=0mm] (scl) {semiconductor material};

    \path[<->]       
        (PT) edge (EL)
        (EL) edge (PN)
    ;

    \begin{pgfonlayer}{background}
        \node [semic_background,
                    fit=(EL) (PN) (scl)] {};
    \end{pgfonlayer}
\end{tikzpicture}
\caption{Sketch of the considered multi-physics system setup. If only the photon intensity is of interest, the degrees of freedom among the semiconductor crystal can be distinguished between direct and indirect coupling.
}
\label{fig:system}
\end{figure}

Traditionally, analytical approximations are applied to parts of the system, to reduce the complexity of the problem and achieve reasonable computational costs. As an example, we consider a semiconductor laser, where the photon intensity in the optical cavity interacts with a coupled electron-phonon gas of the gain medium (see \cref{fig:system}). In the literature, the momentum resolved electron and phonon dynamics are typically reduced by fixing the temperature of the phonons (bath approximation) and treating the electrons in the thermal approximation. Hence, their dynamics are calculated in the relaxation-time approximation for the effective electron temperature\cite{Czycholl}. 
Such approximations compromise the model's predictive power, but have, nonetheless, provided many insights into multi-physics phenomena so far \cite{CHO99, LIN13, STR14a, THU23}.

The successful application of the machine-learning paradigm to technological tasks such as computer vision and natural-language processing \cite{khurana2023natural,bertolini2021machine,sarker2021machine,zhang2019reference,wang2021deep,lee2021short,de2021simulation,zhang2019deep} has motivated researchers to utilize data-driven approaches for scientific discovery \cite{radovic2018machine,brunton2020machine,ourmazed2020,baldi2001bioinformatics,may2021eight,chen2021machine,schmidt2019recent,bedolla2020machine}.
Their application to multi-physics simulation models intends to either improve their accuracy or reduce their computational costs \cite{karniadakis2021physics,peng2021multiscale}.
Successful implementations can be used for efficient parameter studies, multi-objective optimization \cite{wang2023integration,edelen2020machine}, and real-time model based control \cite{xu2020towards}. The different approaches can be distinguished into two classes: The data-driven model can either be designed to entirely emulate the multi-physics system \cite{hennigh2021nvidia,kasim2021building,ladicky2015data,lu2021data,edelen2020machine,bi2023accurate}, or to integrate with system equations \cite{han2020integrating,willard2020integrating,kochkov2021machine,xu2020towards}. Our contribution belongs to the latter class.

In this manuscript, we propose to accelerate the simulation of complex multi-physics systems by approximating the interaction terms among the degrees of freedom, which are not of direct interest, via a data-driven model. Degrees of freedom, which do not directly couple to the observables of interest can even be dropped with their effect implicitly kept in the data-driven model.

Naturally, this approach requires sufficient training data and a suitable data-driven model architecture. Both critically determine the feasibility and success of our approach. 
Given a new problem, an overall improvement is only achieved, if two criteria are met: Firstly, the combined efforts of generating the training data; implementing, training, and testing the data-driven model; and performing the simulations must be computationally less demanding than the direct integration of the system. Secondly, the data-driven approach must provide better accuracy than analytical approximations.
Hence, one must be thoughtful and deliberate with the choice and design of the data-driven model and the generation of the training data.

We demonstrate that our data-driven approach to accelerating multi-physics simulations can outperform a traditional analytic approximation both in terms of accuracy and computational costs on a chosen toy model. The advantage of the data-driven approach becomes especially pronounced if many simulations are required.


To demonstrate our approach, we consider a semiconductor laser. Its dynamics can be separated into two different physical domains, which are represented by the optical resonator with the photon intensity $I$ and the semiconductor crystal with the electron and phonon degrees of freedom \cite{CHO99,HAU04,COL12a}.
Accurate quantitative simulations have been proven difficult, because they involve the self-consistent treatment of both domains, which strongly interact via the stimulated emission/absorption process \cite{CHO99,HAU04}.

For our purpose, we are only interested in the photon intensity at the lasing mode. 
Following our data-driven approach, we therefore construct an approximation model, which only keeps dynamical variables for the photon intensity and the electron states. The phonon states, i.e., the remaining degrees of freedom of the semiconductor crystal, are dropped since they only indirectly couple to the photon intensity.
All the internal interactions of the electron-phonon system are captured by a data-driven model and are then incorporated into the dynamical electron equations.
\Cref{fig:system} visualizes and summarizes our subdivision of the system's degrees of freedom and indicates their couplings.

\section{Methods} \label{sec:models}
To model the semiconductor laser, we devise a toy model, which builds upon our previous work on electron-phonon dynamics \cite{MEI23}, by coupling a photon intensity equation to the electrons via a stimulated emission term \cite{CHO99, LIN15b}.
The dynamics are described by the following equations of motion for the photon intensity $I$ and the electron occupations $f_k$:
\begin{align}
    \ddt I =& -\frac{I}{\tau_\rm{pht}} + \sum_k G(k) D(k) \left[2f_k - 1 \right]I, \label{eq:I}\\
    \ddt f_k =&  -G(k) \left[2f_k - 1 \right]I + \partial_t f_k \vert_\rm{col}.\label{eq:f_k}
\end{align}
The first term in \cref{eq:I} describes the photon lifetime $\tau_\rm{pht}$ and the second term the interaction with the electrons via stimulated emission, where $G(k)$ denotes the $k$-dependent gain function, $D(k)$ the number of states at discretized $k$, and $[2f_k -1]$ the inversion. Note that the sum runs over the modulus $k$ of the two-dimensional momentum $\mathbf{k}$, which yields $D(k) = k\Delta k / 2 \pi$ with the discretization step $\Delta k$. The gain function is given by
\begin{align}
    G(k) = \frac{G_\rm{lin}}{\pi \gamma_\rm{ph} } \sech{\left(\frac{2\epsilon_k - \epsilon_\rm{ph}}{\gamma_\rm{ph}}\right)}
\end{align}
with the linear gain coefficient $G_\rm{lin}$, the homogeneous linewidth of the transitions $\gamma_\rm{ph}$, the single particle kinetic energy $\epsilon_k$, and the lasing-mode photon energy $\epsilon_\rm{ph}$ with respect to the bandgap. The factor of two accounts for the symmetric valance band. The homogeneous linewidth is modeled by a hyperbolic secant to ensure properly decaying tails \cite{CHO99}.
The first term in \cref{eq:f_k} describes the interaction with the photon intensity via stimulated emission and the second term describes the internal processes of the semiconductor material, where $\partial_t f_k \vert_\rm{col}$ summarizes all considered collision integrals. 

\begin{table}[htbp]
\centering
 \caption{Laser parameters}
 \begin{tabular}{c|c|c|c}
   \hline
 $G_\rm{lin}$ & 0.04\,nm$^2$eV/fs & $\tau_\rm{pht}$ & 200\,fs \\  
 $\epsilon_\rm{ph}$ & 0.035\,eV & $\gamma_\rm{ph}$ & 0.01\,eV\\
 \end{tabular}\label{tab:laser}
\end{table}
We choose laser parameters, which are inspired by a nanolaser \cite{THU23, ROO21} and are summarized in \cref{tab:laser}. The characteristically large optical confinement factor and short photon lifetime yield photon intensity dynamics on a time scale similar to the electron dynamics \cite{THU23}. Hence, this setup relies on the proper modeling of the latter and likely highlights any shortcomings.

Our microscopic model for the collision term includes the interaction with two optical and two acoustic phonon branches and omits electron-electron interaction. 
Hence, the microscopic evaluation of the collision term complements the laser model with dynamical equations for the phonon occupation numbers $n_q^\alpha$, where $q$ denotes the crystal momentum and $\alpha$ the branch.
A detailed description is presented in \cref{sec:eom}.
The microscopic evaluation of the collision term
\begin{align}
    \partial_t f_k \vert_\rm{col}^\rm{micro} \left\{ f_k, n_q^\alpha \right\} \label{eq:microscopic_model}
\end{align}
is typically expensive, which in turn drives the total computational costs of the simulation. Hence, one utilizes approximations for the collision term, to achieve reasonable computation times. 

With our data-driven approach, we replace the microscopic evaluation of the collision term with a trained data-driven model
\begin{align}
    \partial_t f_k \vert_\rm{col}^\rm{data} \left\{ f_k(t), f_k(t-\Delta t), \dots, f_k(t-(\ell-1)\Delta t) \right\} \label{eq:data-driven_model}
\end{align}
where, instead of the phonon occupation numbers $n_q^\alpha$, a delay embedding of the electron occupations $f_k$ with the past $\ell$ states is implemented. This yields delay-differential equations and can be thought of as a Takens embedding \cite{TAK81}.
For our purpose, we compose a delay-embedded regressive reduced-order model (derrom) via the package \cite{MEI23}, which we built and maintain ourselves.
In short, derrom first projects the past $\ell$ system states into a reduced dimensionality (order) latent space, which is constructed from the first $r$ left singular vectors of the training data.
This step is designed to retain the dominant patterns of the input data and remove redundant and irrelevant information. 
Next, the latent space features are normalized to appropriate magnitudes and thereafter nonlinearly transformed, to yield derrom's feature vector. The nonlinear transformation is constructed via $L$ hyperbolic tangent activation functions with random weights and biases. Lastly, the regression step is taken via a linear map $\mathbf{W}$. Both the dimensionality reduction and the regression step are trained with the supervised learning paradigm, where we minimize the Frobenius norm of the residual error between the training data and the corresponding reconstruction/prediction. More details on the delay-embedded regressive reduced-order model are presented in \cref{sec:derrom}.

To assess the utility of our proposed data-driven approach, we implement a two-temperature approximation (tta) for the collision term \cite{Czycholl}. This approximation assumes a steady state distribution for electrons and phonons at all times. Hence, the electron and phonon occupations are then described by effective electron and phonon temperature instead of a fully momentum resolved distribution function \cite{Czycholl}
Consequently, the electron dynamics can then be described as an exponential relaxation towards a Fermi-Dirac distribution with the phonon temperature $T_\rm{ph}$. The phonon temperature $T_\rm{ph}$ itself is included as another dynamical quantity, whose relaxation is driven by the difference to the quasi-equilibrium temperature of the electron distribution.
The two-temperature approximation
\begin{align}
    \partial_t f_k \vert_\rm{col}^\rm{tta} \left\{ f_k, T_\rm{ph} \right\} \label{eq:tta_model}
\end{align}
thus reduces the computational costs, but has the major caveat that it is only valid for electron distributions, which are sufficiently close to an equilibrium state.
Such relaxation-time approximations are, nonetheless, often used for the simulation of laser dynamics \cite{CHO99,LIN13,KIL14, MEI19, MEI22,THU23} due to a lack of viable alternatives.
A detailed description of the two-temperature approximation and an example relaxation are presented in \cref{sec:tta}.

\section{Results} \label{sec:results}

\subsection{Test Scenario Setup}

To study the applicability of the data-driven approach, we design a test scenario, in which we are interested in the laser's response to being optically pumped by a short pulse far above the band edge. Specifically, we already know that the laser emits a pulse if the pump energy is sufficient and we want to study how the pulse peak power and the pulse position (the delay after the pump pulse) depend on the power, spectral position, and spectral width of the pump pulse. The pump pulse is implemented as an initial electron state with a Gaussian distribution, where the maximum reflects the power, the mean the center frequency, and the standard deviation the spectral width of the pump laser.

\begin{figure}[htbp]
\centering
\includegraphics[width=\mywidth \linewidth]{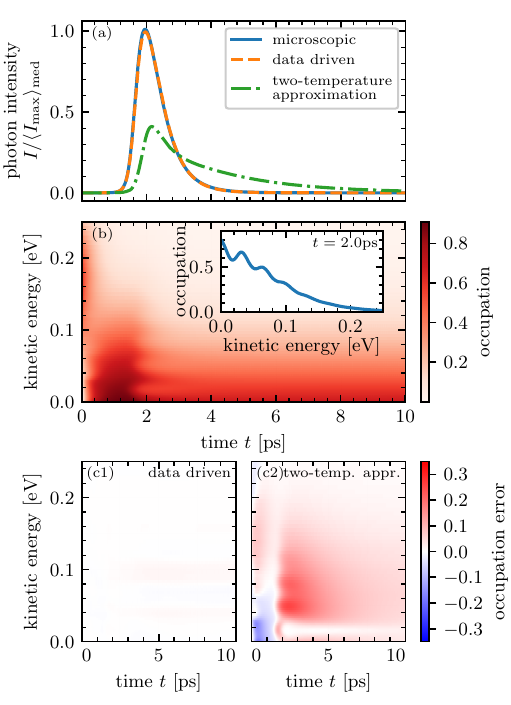}
\caption{Demonstration of the laser dynamics and approximation methods for representative initial conditions.
(a) Photon intensity as obtained from the microscopic model (solid blue), data-driven model (dashed orange), and the two-temperature approximation (dashed-dotted green) normalized to the median pulse maximum of the training/testing data set. (b) Color-coded electron occupation as function of the single-particle energy and time. The inset shows the distribution at $t=2.0\u{ps}$. (c1) and (c2) color-coded electron occupation errors of the data-driven model and the two-temperature approximation, respectively.
}
\label{fig:dynamics_approx_example}
\end{figure}

An example trajectory is presented in \cref{fig:dynamics_approx_example}, where the blue line in (a) shows the normalized photon intensity and (b) shows the conduction-band electron distribution as a function of the single-particle energy. The additional plots and panels regarding the different approximations will be discussed later on. 
After an initial lag of $\approx 1.2\u{ps}$, the photon intensity $I$ rises to form a pulse which peaks around $\approx 2.0\u{ps}$ and then quickly decays again. Much of the observed photon intensity dynamics can be explained by the electron dynamics. Initially, the electrons, which have been excited by the pump with a center energy of $0.160\u{eV}$, decay towards their quasi-Fermi distribution via the electron-phonon interaction. This process supplies the gain to the lasing mode, which is located at $0.0175\u{eV}$ and thus causes some of the observed lag. Once the laser has turned on, the photon intensity clearly burns a spectral hole into the electron distribution. Moreover, a periodic pattern appears with a period of $30\u{meV}$, which corresponds to the energy of the optical phonons. Hence, the refilling of the initial spectral hole via the interaction with optical phonons causes characteristic secondary spectral holes in the electron distribution. The generated pattern is highlighted in the inset in \cref{fig:dynamics_approx_example}\,(b). Once the available gain is used up, the laser pulse quickly dies and the electron distribution approaches a Fermi distribution.
From that example, we can deduce that the details of the electron dynamics, i.e., the collision term, strongly determine the temporal position and maximum value of the photon intensity pulse. Therefore, we score candidate approximations of the collision term with respect to their ability to achieve small errors of those two quantities 
(details in \cref{sec:benchmarking}).

\subsection{Training-Data Generation and Training}

Since, we want to evaluate both the reduction of the simulation time and the accuracy of our data-driven approach, we set ourselves the goal of simulating 1000 trajectories of $10\u{ps}$ length with varying pumping conditions. For that purpose, we construct 1000 initial conditions, where we draw the maximum, mean, and standard deviation of the Gaussian initial conditions from the uniform distributions $\mathcal{U}_\rm{max}(0.5,0.99)$, $\mathcal{U}_\rm{mean}(0.1\u{eV},0.2\u{eV})$, and $\mathcal{U}_\rm{std}(0.04\u{eV},0.06\u{eV})$, respectively.

To operate the data-driven approach, we first require sufficient training data. Since, we a priori don't know the required number of training trajectories, we implement the following strategy: We first set a quantitative accuracy target, which is to be evaluated via a 10-fold cross-validation procedure on the training data (see \cref{sec:benchmarking}).
For our purpose, we define the accuracy measure to be the sum of the 90ths percentiles of the normalized absolute pulse maximum errors $\Delta I_\rm{max}$ and the normalized absolute pulse maximum position errors $\Delta t_{I_\rm{max}}$, i.e.,
\begin{align}
    A = P_{90} \left\{\frac{|\Delta I_\rm{max}|}{\langle I_\rm{max} \rangle_\rm{md}} \right\} 
    + P_{90} \left\{ \frac{|\Delta t_{I_\rm{max}}|}{\langle t_{I_\rm{max}} \rangle_\rm{md}} \right\} \label{eq:accuracy_metric}
\end{align}
and require it to be smaller than $0.2$. Note that we specifically use robust statistics to mitigate the effect of outliers.

In the next step, we start simulating trajectories using the microscopic model in quasi-logarithmic steps, i.e., $10,20,50,100,...$. After each step, we set out to obtain an optimal data-driven model via a heuristic hyperparameter optimization. Based on our previous experience with the electron-phonon system \cite{MEI23}, we keep a fixed delay embedding of $\ell = 2$ and a fixed number of nonlinear nodes $L = 1000$ and optimize the number of reduced dimensions $d_r$ and the regularization strength $\alpha$ via a simple grid search.
This procedure can be assigned to the class of exploration-labeling-training (ELT) algorithms \cite{han2020integrating, zhang2018reinforced}, whereat the exploration is performed via the randomly drawn initial conditions.
Once such a model meets the accuracy target, it can be used for the further simulation of the remaining trajectories.

\begin{figure}[htbp]
\centering
\includegraphics[width=\mywidth \linewidth]{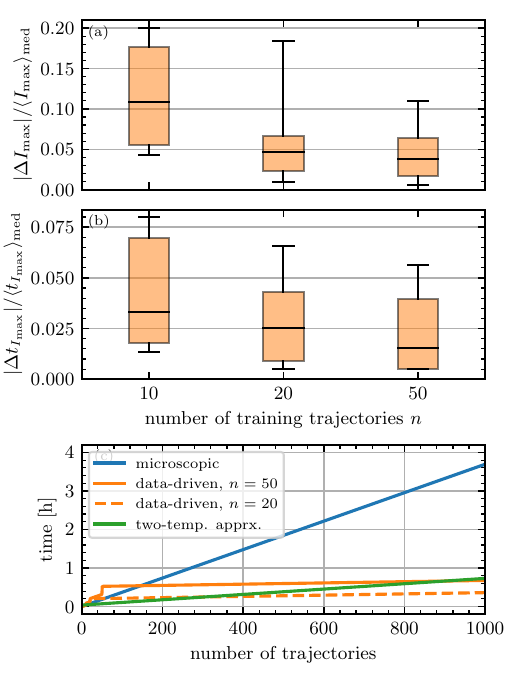}
\caption{Simulation of 1000 trajectories. Cross-validated training set error of the pulse maximum (a) and the pulse maximum position (b) as function of the number of training trajectories $n$ for the best performing model. Only the modulus of the deviations is considered and normalization is performed with respect to the median pulse maximum and median pulse maximum position. The whiskers denote the 10th and 90th percentile.
(c) Simulation time as a function of the number of trajectories. Blue denotes the microscopic model, orange denotes the data-driven model with $n=50$ (solid) and $n=20$ (dashed) training trajectories, and green denotes the two-temperature approximation.
}
\label{fig:training_set_and_time}
\end{figure}

The accuracy results of this strategy are presented in \cref{fig:training_set_and_time}. Panels (a) and (b) show the training set cross-validated errors for the normalized absolute pulse maximum error and the normalized absolute pulse maximum position error as a function of the number training trajectories $n$. In all cases, the best performing model is shown. In both plots, the horizontal black bar denotes the median, the box the interquartile range, and the whiskers the 10th and 90th percentile. Hence, the sum of the top whiskers represents the accuracy metric \cref{eq:accuracy_metric}. For an increasing number of training trajectories, the observed errors decrease, i.e., all indicators shift to smaller values. We find that the accuracy target \cref{eq:accuracy_metric} is met for $n=50$ training trajectories with a value of $A_{n=50} \approx 0.16$. 
Thus, we can now use the obtained data-driven model for the further simulation of trajectories.
Before moving on, however, we would like to stress, that choosing a suitable accuracy target is the human operator's responsibility. Our choice is motivated by achieving a competitive performance with respect to accuracy and overall simulation time. For comparison, the two-temperature approximation only achieves an accuracy of $A_\rm{tta} \approx 1.5$ for the considered 50 training trajectories.

\subsection{Simulation-Time Evaluation}

For the evaluation of the computation time, we simulate all 1000 trajectories using the microscopic model, the data-driven model with the outlined strategy, and the two-temperature approximation. The setup of the latter is described in \cref{sec:tta}. Details on the implementation are given in \cref{sec:implementation}. \Cref{fig:training_set_and_time}\,(c) plots the total simulation time as a function of the number of trajectories. For the microscopic model (blue line), the simulation time increases linearly and amounts to $\approx 3.7\u{h}$ for all trajectories. This corresponds to an average of $\approx 13.3\u{s}$ per trajectory. Similarly, the time for the two-temperature approximation (green solid line) also increases linearly, but only amounts to $\approx 0.73\u{h}$ for all trajectories, which corresponds to $\approx 2.5\u{s}$ per trajectory.

The data-driven approach (orange solid line), on the other hand, shows a distinctive two-stage behavior, which is due to the initial simulation of training data and model hyperparameter optimization. However, once a sufficient model is obtained, it only takes $\approx 0.58\u{s}$ on average to simulate a trajectory. The total simulation time for all 1000 trajectories then amounts to $\approx 0.68\u{h}$. The crossing point with the microscopic model is found at $n=147$ trajectories and with the two-temperature approximation at $n=900$ trajectories. 

At this point, we also want to highlight that the position of crossing points is determined by the initial costs of obtaining the data-driven model. E.g., if we set the accuracy target \cref{eq:accuracy_metric} to $0.25$ instead of of $0.2$, $n=20$ training trajectories are sufficient. The corresponding simulation time is shown via the orange dashed line. For that case, the time to obtain all 1000 trajectories now only amounts to $\approx 0.36\u{h}$ and the crossing point with the microscopic model is found at $n=57$ and with the two-temperature approximation at $n=305$.

In the limit of large numbers of trajectories, i.e., if the initial costs of obtaining the data-driven model can be neglected, the data-driven approach offers an acceleration factor of $\approx 20$ compared to the microscopic model. The two-temperature approximation, on the other hand, only offers a factor of $\approx 4$. While the latter is deceptively easy to write down, its implementation requires to fit a Fermi-function to the electron distribution in each step, which drives the computational costs. The data-driven model, on the other side, only performs a few vector and matrix operations, and one nonlinear transformation, which turn out to be relatively inexpensive. Thus, the data-driven approach also exhibits a performance advantage over the analytic approximation.

\subsection{Accuracy Evaluation}

\begin{figure}[htbp]
\centering
\includegraphics[width=\mywidth \linewidth]{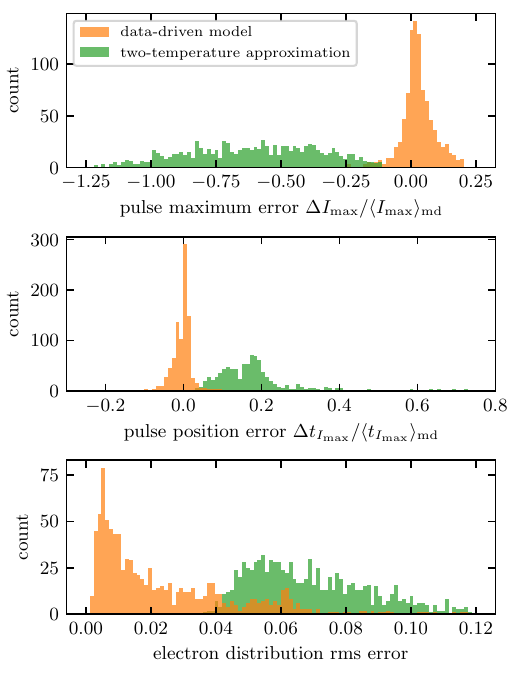}
\caption{Histograms of the pulse maximum error (a), the pulse position error (b) and the electron distribution error (c). Orange denotes the data-driven model and green the two-temperature approximation. The former two are normalized by the median of the pulse maximum and pulse maximum position. The latter amounts to $\langle t_{I_{\rm{max}}} \rangle_{\rm{md}} = 1.93$\,ns.
}
\label{fig:histograms}
\end{figure}

Having benchmarked the simulation time performances, we now continue with the detailed evaluation of the simulation accuracy. Implicitly, we have already done that via the training error in \cref{fig:training_set_and_time} and the accuracy target \cref{eq:accuracy_metric}. At this point, however, we want to strictly separate the training and testing data. For that purpose, we use the optimal data-driven model trained with $n=50$ trajectories and compare its approximation of the 950 trajectories of our test scenario to the trajectories computed with the two-temperature approximation with respect to ground truth produced by the microscopic model. 

To start off with, we would first like to exemplify both approximation methods via the representative trajectory show in \cref{fig:dynamics_approx_example}. In panel (a), the data-driven model is denoted by the orange dashed line and the two-temperature approximation by the green dash-dotted line. We find that the data-driven model beautifully tracks the microscopic model with the pulse maximum only slightly to small and slightly shifted to later times. The two-temperature approximation, on the other hand, only gets the rough qualitative feature right: The pulse maximum is underestimated by more than a factor of two, the pulse is shifted by $\approx 0.5\u{ps}$ to later times, and the trailing edge decays much slower.

The apparent different behavior of the two approximations can be explained by their respective electron distribution dynamics, which are presented in \cref{fig:dynamics_approx_example}\,(c) in terms of the occupation error $f_{\epsilon_k}^\rm{apprx}(t) - f_{\epsilon_k}^\rm{micro}(t)$ with respect to the microscopic simulation shown in (b). The occupation errors produced by the data-driven model \cref{fig:dynamics_approx_example}\,(c1) are hardly visible on the common color scale, since they are about one magnitude smaller than those of the two-temperature approximation \cref{fig:dynamics_approx_example}\,(c2). Moreover, the two-temperature approximation misses important qualitative features of the microscopic model. Firstly, the initial relaxation is to slow and thus produces the blue area in the bottom left of the figure. Secondly, it does not exhibit the characteristic periodic structure produced by interaction with the optical phonons. Lastly, the refilling of the spectral hole, which is caused by the lasing process, also occurs to slowly and thus leaves the pronounced red region at energies above the lasing mode (as identified by the white horizontal stripe for $t > 2\u{ps}$). 

Shortcomings of the two-temperature approximation are to be expected, since the appearing electron distributions are far away from a quasi equilibrium. However, it is important to note, that the quantitative impact onto the photon intensity dynamics cannot be known a priori. Hence, a responsible utilization of the two-temperature approximation should include a quantitative estimation of the potential errors.

To provide a broad picture of the approximation accuracy of the 950 test trajectories, \cref{fig:histograms} shows histograms of the normalized pulse maximum error (a), the normalized pulse maximum position error (b), and the electron distribution rms error (c) for both the data-driven model (orange) and the two-temperature approximation (green).

Taking a look at the pulse maximum error first, we find that the data-driven model produces deviations that are almost centered above zero and hardly exceed $\pm0.2$. The two-temperature approximation, on the other hand, exclusively underestimates the true maximum and generates normalized errors up to $-1.25$ with a median of $\approx -0.6$.

Moving on to the pulse maximum position error, we find that the data-driven model produces errors, which are approximately centered above zero and do not exceed $\pm 0.1$. This time, the two-temperature approximation exclusively overestimates the pulse maximum positions and thus produces normalized errors with a median of $\approx 0.17$ and outliers up to $0.72$.

We can therefore conclude that the two-temperature approximation systematically underestimates the pulse maximum and overestimates the pulse position, while the data-driven model rather suffers from (much smaller) random errors. We attribute the behavior of the two-temperature approximation to the insufficient description of the electron dynamics for distributions for away from a quasi-equilibrium, as discussed previously. 

Hence, we lastly take a look at the electron distribution rms error, which by definition is positive. The data-driven model exhibits a median error of $\approx 0.015$ and the two-temperature approximation a median error of $\approx 0.066$. Note, that the rms error is not specific to the relevant physical features of the electron distribution, e.g., the gain at the lasing mode, and is therefore not an optimal predictor for the pulse accuracy. Nonetheless, it indicates the much better performance of the data-driven method over the two-temperature approximation.

With those results, we conclude that the example trajectories presented in \cref{fig:dynamics_approx_example} are representative both of the data-driven approach and the two-temperature approximation. Both the example and the statistics demonstrate the much better accuracy of the data-driven approach when compared to the two-temperature approximation.
This can also be expressed in terms of the accuracy metric \cref{eq:accuracy_metric}, which yields a value of $A_\rm{tta} \approx 1.38$ for the two-temperature approximation and a value of $A_{n=50} \approx 0.16$ for selected data-driven model on the 950 test trajectories.

\section{Discussion and Conclusions} \label{sec:discussion}

In this work, we have proposed a data-driven approach to accelerate multi-physics simulations and exemplarily demonstrated its application to a semiconductor laser model.
The primary goal was to reduce computational costs without compromising accuracy compared to traditional analytical approximations. In our test scenario, the results establish a substantial advantage of the data-driven method over the two-temperature approximation in both computational efficiency and accuracy.

While demonstrating promising results, this study also has certain limitations. 
Naturally, the quantitative results regarding the accuracy and computation time are specific to the considered system, the mathematical model, the test scenario, and the implementation. 
However, we believe that the presented case study does not represent a best case scenario for data-driven approaches and that the potential performance gains may be much greater in other setups. This is because our toy semiconductor laser model only considers electron-phonon interactions in the collision term and no electron-electron interactions. The latter, however, are numerically more expensive to evaluate \cite{CHO99} and their inclusion would therefore give further advantage to the data-driven approach. This would become especially relevant if a finer discretization was required due to the scaling behavior of the microscopic evaluation (quadratic) and the data-driven model (linear) with the number discretization points.

Moreover, we showed that the data-driven approach only becomes feasible if a certain minimum number of trajectories is required due to the training costs of the data-driven model. However, it is easy to think of cases, where one would need to simulate many more than the 1000 trajectories of our test scenario. For a parameter study, 
a three-dimensional grid with 1000 samples only corresponds to ten sample values per parameter. Adding further parameters or increasing the sample density then quickly drives up the number of samples. If one where to include stochastic effects, e.g., spontaneous emission to our laser, each parameter combination would require multiple trajectory realizations to obtain proper statistics. Hence, the initial training costs can become negligible in many potential cases. 

Both the microscopic evaluation and the data-driven model can benefit from a more efficient implementation. This may include the exact mathematical formulation, the simulation code, and in the case of the data-driven approach, the structure of the model and its training. 
The latter thus offers further options for a more efficient implementation and thus performance improvements. It thereby also puts the burden of doing so on the human operator. This stresses the importance of the careful construction and training of the data-driven model. 
Nonetheless, we are convinced that this task is manageable at present, since our case study already demonstrated excellent results without extensive model design experimentation and optimization.
We expect future research to further establish practical guidelines for the successful design and training of data-driven approximation models.

One promising future research path is the integration of physical knowledge into the data-driven model. This could be done via the structure of the model itself, a combination with a knowledge-based approximation, or the training procedure \cite{willard2020integrating}. 
In our considered case of the semiconductor laser, the proper collision term only redistributes charge-carriers and thus induces no net change of the charge-carrier density. At present, this physical knowledge is not included in the model and therefore small deviations can be observed. A straight-forward option to fixing this, could be the penalization of such deviations during the training via the cost function.



In conclusion, the data-driven approach presented in this study holds significant promise for accelerating multi-physics simulations in various fields, including but not limited to semiconductor lasers. 
Its superior performance in computational efficiency and accuracy, as compared to traditional approximations, may open avenues for further exploration and application in diverse scientific domains and facilitate rapid prototyping and optimization of complex systems.


\section{Acknowledgements}

We acknowledge fruitful discussions with 
Dominik Christiansen (TU Berlin).\\
This work was funded by the Deutsche Forschungsgemeinschaft (DFG) through Project SE 3098/1, Project No. 432266622 (M.S.), SFB 951, Project No. 182087777 (A.K.).

\section{Author contributions}
S.M. and M.S. initiated and conceptualized the work. M.S. and A.K. worked on the microscopic derivation and the numerical implementation of electron-phonon scattering and parameter search. S.M. implemented the laser model and the data-driven model; performed the simulations and numerical experiments; and drafted the manuscript. All authors discussed and edited the manuscript.

\section{Data Availability}
The data generated in this work can be generated by running the publicly available code as described in the code availability statement.

\section{Code Availability}
The simulation code and the regression code is available on GitHub under MIT license (\url{https://github.com/stmeinecke/derrom}) 

\appendix

\section{Microscopic Electron-Phonon Equations} \label{sec:eom}

We have described the coupled electron-phonon dynamics in our previous publication \cite{MEI23} and only present a brief summary here. The equations of motion for the electron occupations $f_\mathbf{k}$ and phonon occupations $n_\mathbf{q}^\alpha$ are derived from a Hamiltonian via the Heisenberg equation. The electron dispersion is treated in the parabolic approximation, the acoustic phonon dispersion in the Debye approximation, and the optical phonon dispersion in the Einstein approximation, with parameters from ab-initio calculations\cite{Kormanyos2015,Li2013}. The occurring hierarchy problem is addressed using a correlation expansion and a second-order Born-Markov approximation\cite{butscher2007hot}. 
The resulting coupled electron-phonon Boltzmann scattering equations then read
\begin{widetext}
\begin{align}
\partial_t f_\mathbf{k}\vert_\rm{col}^\rm{micro} &= \frac{2 \pi }{\hbar} \sum_{\mathbf{k'},\alpha,\pm} |g_{\mathbf{k} - \mathbf{k'}}|^2 \left(\frac{1}{2} \pm \frac{1}{2} + n_\mathbf{\mathbf{k} - \mathbf{k'}}^\alpha \right) \left( 1 - f_\mathbf{k} \right) f_\mathbf{k'} \delta (\epsilon_\mathbf{k} - \epsilon_\mathbf{k'} \pm \hbar \omega^\alpha_\mathbf{k - k'})\nonumber \\
&- \frac{2 \pi }{\hbar} \sum_{\mathbf{k'},\alpha,\pm} |g_{\mathbf{k} - \mathbf{k'}}|^2 \left(\frac{1}{2} \pm \frac{1}{2} + n_\mathbf{\mathbf{k} - \mathbf{k'}}^\alpha \right) \left( 1 - f_\mathbf{k'} \right) f_\mathbf{k} \delta (\epsilon_\mathbf{k} - \epsilon_\mathbf{k'} \mp \hbar \omega^\alpha_\mathbf{k - k'})\label{eq:elec} \\
\partial_t n_\mathbf{q}^\alpha &= \frac{2 \pi }{\hbar} |g_\mathbf{q}^\alpha|^2 \sum_\mathbf{k}  \left( 1 - f_\mathbf{k} \right) f_\mathbf{k+q} \left( 1 + n_\mathbf{q} \right) \delta (\epsilon_\mathbf{k} - \epsilon_\mathbf{k+q} + \hbar \omega^\alpha_\mathbf{q})\nonumber \\
&- \frac{2 \pi }{\hbar} |g_\mathbf{q}^\alpha|^2 \sum_\mathbf{k}  \left( 1 - f_\mathbf{k+q} \right) f_\mathbf{k} n_\mathbf{q}  \delta (\epsilon_\mathbf{k} - \epsilon_\mathbf{k+q} - \hbar \omega^\alpha_\mathbf{q}).\label{eq:phon}
\end{align}
\end{widetext}

The appearing electron coupling elements $g_\mathbf{k}$ are treated in the effective deformation potential approximation \onlinecite{Li2013}
\begin{equation}
g_q^{i} = \sqrt{\frac{\hbar}{2 \rho \Omega^i A}} V_q,
\end{equation}
with the effective mass density of the unit cell $\rho$ and the semiconductor area $A$. The effective deformation potential reads $V_q=D_1 q$ for acoustic phonon coupling, and $V_q=D_0$ for optical phonons. We evaluate the system for two-dimensional MoSe2 \cite{Li2013,Jin2014,Steinhoff2014,perea2019exciton,selig2019ultrafast} and take the parameters from reference \onlinecite{Jin2014} and list them in \cref{tab_e_phon}.

\begin{table}[h!]
\centering
 \caption{Material parameters.}
 \begin{tabular}{c|c|c}
   \hline
 $m_e$/$m_0$ & 0.5 & \onlinecite{Kormanyos2015} \\  
 $\bar \omega^o$/meV & 36 & \onlinecite{Jin2014}\\
 $c_{LA}$/(nm/fs) & 4.1 & \onlinecite{Jin2014}\\
 $D^{a}_1$/eV & 3.4 & \onlinecite{Jin2014}\\
 $D^{o}_0$/eV nm$^{-1}$ & 52 & \onlinecite{Jin2014}\\
 \end{tabular}\label{tab_e_phon}
\end{table}

\section{Delay Embedded Regressive Reduced Order Model} \label{sec:derrom}

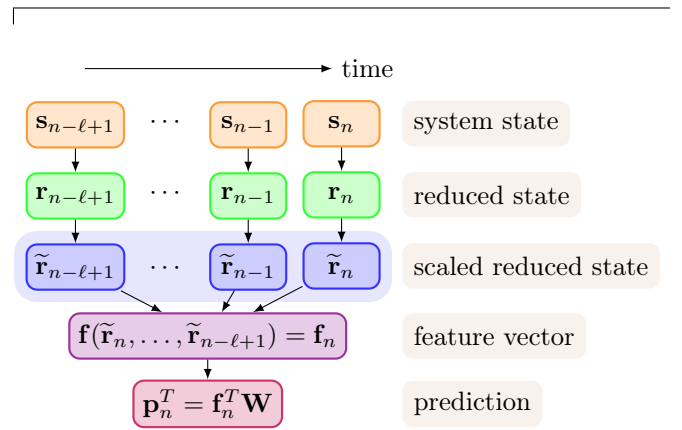
\begin{figure}[htbp]
\centering
\tikzstyle{state}=[rectangle,
    thick,
    minimum height=0.6cm,
    minimum width=1.0cm,
    draw=orange!80,
    fill=orange!20,
    rounded corners]
    
\tikzstyle{rstate}=[rectangle,
    thick,
    minimum height=0.6cm,
    minimum width=1.0cm,
    draw=green!80,
    fill=green!20,
    rounded corners]
    
\tikzstyle{srstate}=[rectangle,
    thick,
    minimum height=0.6cm,
    minimum width=1.0cm,
    draw=blue!80,
    fill=blue!20,
    rounded corners]
    
\tikzstyle{dummy}=[rectangle,
    minimum height=0.6cm,
    minimum width=0.1cm]

\tikzstyle{tdummy}=[rectangle,
    minimum height=0.1cm,
    minimum width=0.1cm]

\tikzstyle{nltstate}=[rectangle,
    thick,
    minimum height=0.6cm,
    minimum width=1.2cm,
    draw=violet!80,
    fill=violet!20,
    rounded corners]
    
\tikzstyle{pred}=[rectangle,
    thick,
    minimum height=0.6cm,
    minimum width=0.6cm,
    draw=purple!80,
    fill=purple!20,
    rounded corners]
    
\tikzstyle{line} = [draw, -latex']

\tikzstyle{desc}=[rectangle,
    minimum height=0.6cm,
    minimum width=0.6cm,
    fill=brown!10,
    rounded corners]

\tikzstyle{background}=[rectangle,
    fill=gray!10,
    inner sep=0.15cm,
    rounded corners=3mm]
    
\tikzstyle{background_VAR}=[rectangle,
    fill=blue!10,
    inner sep=0.15cm,
    rounded corners=3mm]

\begin{tikzpicture}[>=latex]

  \matrix (mtrx) [row sep=0.3cm, column sep=0.2cm, matrix of nodes, nodes in empty cells] {
    \node (d1_1) [tdummy] {}; & & & \node (d1_4) [tdummy] {};\\
    \node (s_n_1) [state]{$\mathbf{s}_{n-\ell+1}$}; & \node (dots1) {$\cdots$};  & \node (s_n_2) [state]{$\mathbf{s}_{n-1}$}; & \node (s_n_3) [state]{$\mathbf{s}_{n}$}; \\
    \node (r_n_1) [rstate]{$\mathbf{r}_{n-\ell+1}$}; & \node (dots2) {$\cdots$}; & \node (r_n_2) [rstate]{$\mathbf{r}_{n-1}$}; & \node (r_n_3) [rstate]{$\mathbf{r}_{n}$}; \\
    \node (sr_n_1) [srstate]{$\mathbf{\widetilde{r}}_{n-\ell+1}$};  & \node (dots3) {$\cdots$};  & \node (sr_n_2) [srstate]{$\mathbf{\widetilde{r}}_{n-1}$}; & \node (sr_n_3) [srstate]{$\mathbf{\widetilde{r}}_{n}$};\\
    \node (d4) [dummy] {}; & & & \\
    \node (d5) [dummy] {}; & & & \\
    };
    
    \node (nltpos) [fit=(mtrx-5-2) (mtrx-5-3)]{};
    \node (ppos) [fit=(mtrx-6-2) (mtrx-6-3)]{};
    
    \draw (nltpos) node (nlt) [nltstate]{$\mathbf{f}(\mathbf{\widetilde{r}}_{n},\dots, \mathbf{\widetilde{r}}_{n-\ell+1}) = \mathbf{f}_n$};
    
    \draw (ppos) node (p_n) [pred]{$\mathbf{p}_{n}^T = \mathbf{f}_n^T \mathbf{W}$};
    
    \draw (s_n_3) node[desc, right, inner sep=1.5mm, outer sep=8mm] {system state};
    \draw (r_n_3) node[desc, right, inner sep=1.5mm, outer sep=8mm, align=center] {reduced state};
    \draw (sr_n_3) node[desc, right, inner sep=1.5mm, outer sep=8mm, align=center] {scaled reduced state};
    \draw (mtrx-5-4) node[desc, right, inner sep=1.5mm, outer sep=8mm, align=center] {feature vector};
    \draw (mtrx-6-4) node[desc, right, inner sep=1.5mm, outer sep=8mm, align=center] {prediction};
    
    \path[->]
        (d1_1) edge node[pos=1.0,right] {time} (d1_4)
        
        (s_n_1) edge (r_n_1)
        (s_n_2) edge (r_n_2)
        (s_n_3) edge (r_n_3)

        (r_n_1) edge (sr_n_1)
        (r_n_2) edge (sr_n_2)
        (r_n_3) edge (sr_n_3)
        
        (sr_n_1) edge (nlt)
        (sr_n_2) edge (nlt)
        (sr_n_3) edge (nlt)
        
        (nlt) edge (p_n)
    ;
    
    \begin{pgfonlayer}{background}
        \node [background_VAR,
                    fit=(sr_n_1) (sr_n_3)] {};
    \end{pgfonlayer}
\end{tikzpicture}
\caption{Sketch of the delay-embedded regressive reduced order model (derrom). The past $\ell$ system states $\mathbf{s}_n$ are reduced in their dimensionality $\mathbf{r}_n$, scaled $\mathbf{\widetilde{r}}_{n}$, concatenated and then nonlinearly transformed to yield the feature vector $\mathbf{f}_n$. The prediction $\mathbf{p}_n$ is computed via a linear map with the trained weights $\mathbf{W}$.
}
\label{fig:derrom}
\end{figure}

In this section, we present and discuss the delay-embedded regressive reduced order model (derrom) in detail. The processing and flow of information, coming from the past $\ell$ system states, can be separated into multiple individual steps and is illustrated in \cref{fig:derrom}. In this figure, time moves to the right and information flows to the bottom. 
In the first step, the past $\ell$ (truncated) system states $\mathbf{s}_n$ are collected (orange boxes). If they are truncated, only a subset of the system state variables is considered, e.g., only the electron occupation numbers enter the data-driven model in this work. In the next step, each system state is projected into a reduced order latent space $\mathbf{r}_n$ (green boxes). Next, a scaling of each feature, i.e., element in the reduced states, produces the scaled reduced system states $\mathbf{\widetilde{r}}_{n}$ (blue boxes). Their concatenation (light blue box) represents the linear feature vector of the model. The nonlinear feature vector $\mathbf{f}_n$ (purple box) is then generated via the nonlinear transformation $\mathbf{f}(\cdot)$. Note that the feature scaling is highly relevant, because it affects the nonlinear response of $\mathbf{f}$. Lastly, a linear regression step is carried out via the matrix $\mathbf{W}$ to produce the prediction $\mathbf{p}$.


For the dimensionality reduction stage, we use the left singular vectors of the singular value decomposition (SVD) \cite{GOL13a, BRU22}
\begin{align}
    \mathbf{S}^T = \mathbf{U} \mathbf{\Sigma} \mathbf{V}^T, \label{eq:SVD}
\end{align}
which has proven most efficient in our previous work on the electron-phonon system \cite{MEI23}. The SVD has a rich history in science and engineering \cite{GOL13a, BRU22} and is known across different disciplines as the Karhunen–Lo{\`e}ve transform (KLT) \cite{KAR47,LOE17a}, empirical orthogonal functions \cite{LOR56}, proper orthogonal decomposition (POD) \cite{LUM67}, and canonical correlation analysis \cite{CHE96}.
The $\mathbf{U}$ and $\mathbf{V}$ matrices are unitary and contain the left and right singular vectors of $\mathbf{S}^T$. $\mathbf{\Sigma}$ is a diagonal matrix with the singular values in descending order. Their magnitude naturally organizes the left and right singular vectors according to their share in the reconstruction of the data matrix $\mathbf{S}^T$. Specifically, the Eckart-Young theorem \cite{ECK36,BRU22} states that the best rank-$r$ approximation of a matrix with respect to the Frobenius norm can be achieved via the truncation over the leading $r$ singular values of the SVD. 
To reduce a system state $\mathbf{s}_n$, we compute the SVD of the training data to obtain the basis $\mathbf{U}$. We then use the truncated matrix $\mathbf{U}_{r} = (\mathbf{u}_1,\dots,\mathbf{u}_{r})$
to project $\mathbf{s}_n$ onto the first $r$ left singular vectors $\mathbf{u}_m$ to obtain the reduced state
\begin{align}
    \mathbf{r}_n = (r_1,\dots,r_{r})^T = \mathbf{U}_{r}^T \mathbf{s}_n.
\end{align}
Note, that this dimensionality reduction scheme is entirely data driven. We thus expect its performance to depend on the quality of the training data. To obtain a basis, that is well suited for the task, the training data should thus be representative of the relevant system state subspace.


In the feature scaling stage, we standardize the individual features by subtracting their mean $\mu_m$ and scaling them to unit variance
\begin{align}
    \widetilde{r}_{nm} =  \frac{ r_{nm} - \mu_m }{ \sigma_m }, \label{eq:scaling}
\end{align}
where $\sigma_m$ denotes the standard deviation of the $m$-th feature in the training data.


The delay embedding is achieved by concatenating the past $\ell$ system states $\widetilde{\mathbf{r}}$
\begin{align}
    \mathbf{f}_n^\rm{lin} = \widetilde{\mathbf{r}}_n \oplus \widetilde{\mathbf{r}}_{n-1} \oplus \dots \oplus \widetilde{\mathbf{r}}_{n-\ell+1},
\end{align}
where $\oplus$ denotes the concatenation operator. The result represents the linear feature vector $\mathbf{f}_n^\rm{lin}$ of the data-driven model. For a reduced dimensionality $d_r$ of the state vectors $\widetilde{\mathbf{r}}_n$, the dimension of the linear feature vector $\mathbf{f}_n^\rm{lin}$ thus becomes $d_r \ell$.


For the nonlinear transformation, we construct the feature vector according to
\begin{align}
    \mathbf{f}_n = 1 \oplus \mathbf{f}_n^\rm{lin} \oplus \mathbf{f}^\rm{nl}( \mathbf{f}_n^\rm{lin} ), \label{eq:feature_vector}
\end{align}
where we add a bias term, denoted by the one, and explicitly keep the linear feature vector. The nonlinear features are represented by $\mathbf{f}^\rm{nl}( \mathbf{f}_n^\rm{lin} )$.
In our case, the nonlinear transformation is given by
\begin{align}
    \mathbf{f}^\rm{nl}( \mathbf{f}_n^\rm{lin} ) = \phi \left( \mathbf{W}_\rm{nl} \mathbf{f}_n^\rm{lin} + \boldsymbol{\beta} \right),
\end{align}
where a nonlinear activation function $\phi$ acts on each element of the input vector, which is generated by the weight matrix $\mathbf{W}_\rm{nl} \in \mathbb{R}^{L \times d_f}$, the input (linear feature) vector $\mathbf{f}_n^\rm{lin}$, and the bias vector $\boldsymbol{\beta} \in \mathbb{R}^{L \times d_f}$. 
This nonlinear transformation can be represented by a fully connected feed-forward network with one hidden layer, to which a nonlinear activation function is applied. 
The matrix $\mathbf{W}_\rm{nl}$ defines the number of nonlinear nodes $L$ in the hidden layer via its shape $\mathbb{R}^{L \times d_f}$.
The weights $w_{mn}^\rm{nl}$ are drawn from a normal distribution $\mathcal{N}^{\mathbf{W}}(0,d_f^{-1})$ with zero mean and the variance given be the inverse feature vector dimension $d_f$.
Given the standardization of the features, the scaling of the variance ensures that the expected magnitude of the inputs $u_{nl} = \sum_m w_{lm}^\rm{nl} f^\rm{lin}_{nm}$ is of the order $\approx 1$.
This design has been shown to best harness the $\tanh$ nonlinaerity and produce optimal regression results \cite{AKU15}.
With the same argument, the biases $\beta_m$ are drawn from the uniform distribution $\mathcal{U}^\mathbf{\beta}(-1.0,1.0)$.
If one does not optimize the parameters of the transformation, i.e., $\mathbf{W}_\rm{nl}$ and $\boldsymbol{\beta}$, in the training of the model, this approach is also known as an extreme learning machine (ELM) \cite{HUA04c,HUA06a,HUA11a,HUA06,PYL21}.

Lastly, the regression step is performed via the linear map
\begin{align}
    \mathbf{p}_{n}^T = \mathbf{f}_n^T \mathbf{W},
\end{align}
where $\mathbf{W} \in \mathbb{R}^{d_f \times d_p}$ denotes the regression weight matrix with the dimensions of the feature vector $d_f$ and the prediction $d_p$.

The model is trained using the supervised learning paradigm, i.e., the model parameters are tuned to minimize a loss function of some labeled training data and its corresponding model output.
The training of our proposed model includes the dimensionality reduction stage and the regression stage. For the sake of training simplicity, we train the dimensionality reduction stage and the regression weights separately. I.e., we compute the SVD for the dimensionality reduction stage and then optimize the regression weights $\mathbf{W}$ for a given reduced dimensionality $d_r$.

To facilitate the model training, we first sample $M$ trajectories at $N_M$ discrete times and build the data matrix
\begin{align}
    \mathbf{S} = (\mathbf{s}_1^1,\dots,\mathbf{s}_n^m,\dots,\mathbf{s}_{N_M}^M)^T,    
\end{align}
where the rows correspond to the system state vectors $\mathbf{s}_n^m$ of the $m$-th trajectory at the time $t_n^m$.
We then proceed to construct the feature matrix
\begin{align}
    \mathbf{F} = (\mathbf{f}_1^1,\dots,\mathbf{f}_n^m,\dots,\mathbf{f}_{N_M}^M)^T,   \label{eq:feature_matrix}
\end{align}
where the feature vectors $\mathbf{f}_n^m$ are computed according to \cref{eq:feature_vector} from the system states. For delay embeddings ($\ell > 1$), the trajectories $\mathbf{s}_n^m$ are padded with the trajectories first state $\mathbf{s}_1^m$ to calculate the first feature vectors $\mathbf{f}^m_{n<\ell}$.
Lastly, we construct the target matrix
\begin{align}
    \mathbf{T} = (\mathbf{t}_1^1,\dots,\mathbf{t}_n^m,\dots,\mathbf{t}_{N_M}^M)^T,    
\end{align}
where the vector $\mathbf{t}_n^m$ is to be predicted based on the feature vector $\mathbf{f}_n^m$. In this work, it represents the $k$-resolved collision terms in \cref{eq:f_k}. The regression weights $\mathbf{W}$ are then determined by solving the least-squares problem
\begin{align}
    \argmin_{\mathbf{W}} \left[ \norm{\mathbf{F}\mathbf{W} - \mathbf{T}}_\rm{F}^2 + \alpha \norm{\mathbf{W}}_\rm{F}^2 \right], \label{eq:lossfunction}
\end{align}
where $\norm{\cdot}_\rm{F}$ is the Frobenius norm and $\alpha$ the Tikhonov regularization (ridge) parameter \cite{VOG02,PRE07,BRU22}. A nonzero $\alpha$ penalizes large weights $w_{nm}$ and thereby counteracts overfitting while also ensuring that a solution exists. The solution to this problem reads
\begin{align}
    \mathbf{W} = \left( \mathbf{F}^T \mathbf{F} + \alpha \mathbf{I} \right)^{-1} \mathbf{F}^T \mathbf{T},
\end{align}
where $(\cdot)^{-1}$, depending on the rank of $\left( \mathbf{F}^T \mathbf{F} + \alpha \mathbf{I} \right)$, either denotes the inverse or the Moore–Penrose pseudo inverse.

Apart from the model parameters, which are optimized in the training process, four hyperparameters remain: the delay embedding depth $\ell$, the reduced system state dimension $d_r$, the number of nonlinear nodes $L$, and the regularization parameter $\alpha$. To obtain a well performing model, those must be set to appropriate values, either by systematic hyperparameter optimization schemes or based on educated guesses.

\section{Two-Temperature Approximation} \label{sec:tta}

The two-temperature approximation improves upon a relaxation-time approximation with a fixed phonon-bath temperature by introducing a dynamical phonon temperature $T_\rm{ph}$, which represents the combined quasi-equilibrium temperature of all phonon branches. Hence, the heating or cooling of the semiconductor lattice via the electron-phonon interaction is taken into account, albeit within a relaxation-time approximation: The dynamics of the phonon temperature are driven by its difference to the quasi-equilibrium temperature of the electrons. This modification causes the electrons to relax towards a Fermi distribution with the common equilibrium temperature of both the electron and the phonon system. In the considered setup, the common equilibrium temperature depends upon the initial non-equilibrium electron distribution, which is generated by the pump pulse. The respective equations of motion for the electron occupation numbers and the phonon temperature then read
\begin{align}
    \partial_t f_k \vert_\rm{col}^\rm{tta} =& \frac{1}{\tau_\rm{el}} \left[ F(\epsilon_k, \mu, T_\rm{ph}) - f_k \right], \label{eq:tta1}\\
    \ddt T_\rm{ph} =& \frac{1}{\tau_\rm{ph}} \left[ T_\rm{el} - T_\rm{ph} \right], \label{eq:tta2}
\end{align}
where $F(\cdot)$ denotes the Fermi distribution, $\tau_\rm{el}$ and $\tau_\rm{ph}$ the relaxation time-constants for the electron distribution and the phonon temperature, and $T_\rm{el}$ the quasi-equilibrium temperature of the electrons.

To evaluate the right-hand-side of the equations \ref{eq:tta1} and \ref{eq:tta2}, both the chemical potential $\mu$ and the quasi-equilibrium temperature $T_\rm{el}$ must be calculated dynamically. 
$T_\rm{el}$ is uniquely determined by the electron density and the total electron energy density of the current distribution $\{f_k\}$ and $\mu$ is uniquely determined by the electron density (for a given $T_\rm{ph}$). Hence, both can be obtained by solving the optimization problem of matching the density and total energy density of the quasi-Fermi distribution to the current electron distribution. The optimizations are solved iteratively via Newton's method. Note that this process requires multiple evaluations of the Fermi function, which represents the largest contribution to the computational costs of this approximation.

\begin{figure}[htbp]
\centering
\includegraphics[width=0.6\linewidth]{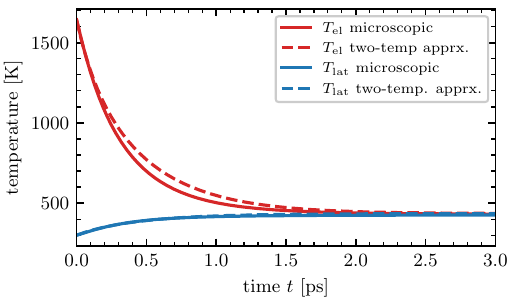}
\caption{Temperature relaxation of the coupled electron-phonon system. The electron system is plotted in red and the phonon system in blue. Solid lines indicate the microscopic model and dashed lines the two-temperature approximation with the time constants $\tau_\rm{el} = 500\u{fs}$ and $\tau_\rm{lat} = 4000\u{fs}$.
}
\label{fig:two_temp_fit}
\end{figure}

To setup the two-temperature approximation, one must also determine the time constants $\tau_\rm{el}$ and $\tau_\rm{ph}$. Since, we have the microscopic model at hand, we treat the time constants as fit parameters, which are to be determined from a sample trajectory. For that purpose, we simulate the electron-phonon system without the photon intensity. The phonons are initialized with a $T=300\u{K}$ Bose-Einstein distribution and the electrons with an out-of-equilibrium Gaussian distribution. The electron-phonon interactions then drives the coupled system towards an equilibrium state, which is characterized by a common temperature.

\Cref{fig:two_temp_fit} plots the quasi-equilibrium temperature of the electrons (red) and the phonons (blue) with solid lines as a function of time. These temperatures are obtained by fitting Fermi/Bose distributions to the electron and phonon distributions as described above. We compute the combined phonon temperature $T_\rm{ph}$ as the mean of the two optical branches. The acoustic branches are neglected, because their much smaller excitation energies lead to temperature dynamics much slower than the considered $10\u{ps}$. For this time interval, we thereby obtain a better approximation for collision term in the electron equation.
For the given parameters, the phonons absorbs energy from the hot electrons and both relax towards the common equilibrium temperature of $T_\rm{eq} \approx 429\u{K}$.

To fit the time constants, we only consider the trajectory for $t>1\u{ps}$, where the system is somewhat close to its equilibrium. $\tau_\rm{el}$ is obtained by fitting an exponential function to the relaxation of the total electron energy density. 
$\tau_\rm{ph}$ is obtained by numerically evaluating the derivative $\tfrac{d}{dt}T_\rm{ph}$ from data and treating \cref{eq:tta2} as an regression equation. This procedure yields $\tau_\rm{el} \approx 424\u{fs}$ and $\tau_\rm{ph}\approx 4333\u{fs}$. However, if we use those values and propagate the same initial conditions with the two-temperature approximation, we underestimate the final equilibrium temperature with $T_\rm{eq} \approx 408\u{K}$. This discrepancy is due to the nonlinearities of the far-out-of equilibrium states at the early stages of the relaxation. To account for that, the manually change time-constants to $\tau_\rm{el} = 500\u{fs}$ and $\tau_\rm{lat} = 4000\u{fs}$ to obtain the proper equilibrium temperature. 
The resulting temperatures for the electrons (red) and the phonons (blue), as obtained from the two-temperature approximation simulation, are plotted in \cref{fig:two_temp_fit} with dashed lines. Besides the final equilibrium temperature, we find a good agreement for the phonon temperature and a decent agreement for the electron temperature. Deviations occur in the initial stages of the relaxation, where the system is still far from its equilibrium.

\section{Benchmarking} \label{sec:benchmarking}

To evaluate the quality of a given approximation (indicated by a hat, sampled at discrete $t_n$), we use the following three metrics: 
The pulse maximum error of the photon intensity $I$
\begin{align}
    \Delta I_\rm{max} = \max_{n} I_n - \max_{n} \hat{I}_n,
\end{align}
the position error of maximum photon intensity $I_\rm{max}$
\begin{align}
    \Delta t_{I_\rm{max}} = \argmax_{t} I - \argmax_{t} \hat{I},
\end{align}
and the root-mean-squared (rms) error of the electron distribution $f$
\begin{align}
    \epsilon_\rm{rms} = \sqrt{ \frac{1}{N} \sum_{n,k} \left( \hat{f}_{nk} - f_{nk} \right)^2 }.
\end{align}

To benchmark a given data-driven model, we use a k-fold cross validation scheme, where we split the training data set (in terms of trajectories) into k equally sized folds. For each fold, we train the model on all the other folds and then score the trajectories contained in the selected fold. This way, we do not mix training and testing data but, nonetheless, obtain a score for each trajectory in the data set. Using the individual error scores, we can then compute the desired statistics, e.g., the median score.

\section{Implementation and Simulation Details} \label{sec:implementation}

For the numerical integration of the laser model and its approximations, we chose a discretization of the electron momentum $k$ with $k_\rm{max} = 2.5\u{nm}^{-1}$ and 100 equidistant points. For the microscopic model, the crystal momentum $q$ follows the same discretization. This choice produces a sufficiently smooth gain curve  (as constructed from the inhomogeneously broadened gain medium) and minimal energy losses in the electron-phonon interaction while remaining computationally tractable.

Both the microscopic model and the two-temperature approximation are then composed of coupled systems of ordinary differential equations (ODEs). To integrate them, we use SciPy's initial value problem solver with the explicit Dormand-Prince adaptive step-size Runge-Kutta method of order 5(4). We note, that increasing the relative and absolute error tolerances does hardly speed up the simulation, but rather destabilizes the integration.
The data-driven approximation, on the other hand, yields a system of delay-differential equations (DDEs) via its delay embedding. Those are outside the scope of the typical ODE solver package. Hence, we implement an explicit second-order Heun method with a constant step-size. This approach has the advantage, that it only evaluates the delayed variables at integer multiples of the step size and does not require any history-array interpolation routine. For our system, this method converges with a step-size $h=5\u{fs}$, which corresponds to half of our sampling interval $\delta t = 10\u{fs}$. 
For a larger output sampling interval, we would advise to implement a constant step-size explicit fourth-order Runge-Kutta method with third-order Hermite interpolation for the history array.

The simulation code for all models in completely implemented in Python and publicly available as stated above.
All simulation were performed on a single CPU, namely a state of the art \emph{Intel(R) Core(TM) i9-10900}. The utilized NumPy and SciPy libraries are allowed utilize all CPU cores.

\bibliography{extracted}

\end{document}